# THE WEAKNESS OF THE SCIENTIFIC ASSESSMENTS:
## A PRAISE OF SILENCE

**José Carlos Bermejo Barrera**
**University of Santiago de Compostela**

### Abstract

This article aims to show the weakness of the current scientific assessments, based on a set of contradictory pseudo-axioms. The six pseudo-axioms are deeply analysed. From the analysis are derived several conclusions. In spite of the serious efforts of the scientists to establish a ranking of honors in order to get funds or to control the academic and higher education institutions, the Science doesn't exists in itself, but different kinds of knowledge. Therefore, the scientists don't control the Science; they are mere experts in a field of knowledge, which could be expressed in many valid ways. This variety determines the existence of many models of academic *curricula*, based on heterogeneous ways of valuation that change along the History.

------

Since some decades in the industrialized world has been developed a system or a series of systems designed to set up a collection of parameters in order to value objectively or even to quantify the progress of the knowledge or of the Science (always starting from the supposition that the knowledge is scientific by definition). Setting up these parameters is very important, because on one hand they are aimed at arranging the funding of the investigation called scientific and on the other hand the foundation of research and teaching institutions and, consequently, the money for the staff of researchers. That determines the professional and living future of a very big group of who we can consider «professionals of knowledge».

We intend to point out below that the main systems of scientific assessment are based on a set of assumptions or pseudo-axioms in contradiction. And we are going to hold also that, if this happens, it is not due to the lack of intelligence of who establishes those systems, but to his ill will or, in other words, to the aim of developing an strategy of fundraising and taking up of the higher education institutions and research centres in favour of certain scientific communities and to the detriment of almost everyone else.

These pseudo-axioms are the following:

1. Only the scientists have the ability of assessing the development of their own subject and the overall system —institutional, political and economic— of production of the knowledge.

2. The production of knowledge is a formalized process, even quantifiable, according to some parameters of universal validity.

3. Just like any process of quantifying, it is essential establish a unit of measure, and that unit of measure must be a publication of any kind.

4. The publications are arranged in order of importance, according to criteria easily objectivable.

5. The publications have an average lifespan that depends on the dynamic of the fast discovering, which characterizes the most of the fields of the scientific research.

6. That dynamic of the fast discovering allows that the Science could do systematically without its past and forecast its own future.

Let`s see the weakness of these pseudo-axioms.

**Weakness of 1:**

A Science could be defined in this way: We call Science to a limited field of the knowledge that must have, first, an object or a group of objects defining its contents; and secondly, a method or a group of methods for examining, analysing and creating principles about those objects.

The objects of the called Natural Sciences —Physics, Chemistry, Biology, Geology...— have a physical existence and are outer to the subject or the subjects studying them —to whom we call «scientific communities»—. In the Formal Sciences —Mathematics, Logic...—, those objects have not physical existence, but we could say that they are outer to the subject and contain a reality of some kind that only can be known through a big effort.

The scientists, or a scientific community, Cc (x) are gifted with the ability to analyse a sector of the reality and, consequently, they can produce principles about that sector of the reality. By knowing a certain kind of objects and having a good grasp of the method of analysis, they can *speak with authority about them,* but only about them. They cannot speak with authority about another kind of objects because they have not the instruments for watching them and are not fluent in the language for speaking about them —in the case of the Physic Science, for example, different branches of the Mathematics—.

If it is true that the scientist has competence only to speak with authority within the limits of his own field, it will be also true that he cannot speak with authority about the overall system of the production of knowledge, because that is not the object of his Science and he doesn't know the proper method to speak about it.

And if this is true about the general system of knowledge, it will be most when dealing with the overall systems —institutional, political and economical—. Because they develop the material and institutional conditions that make feasible the elaboration of the knowledge or of the Science.

We could say that the scientist is the speaker of a language. The speakers of a language have *linguistic competence*, and this allows them to express themselves and to live into the social framework of spreading of that language, but they are not spontaneous grammarians. Of course, they maybe ignore the syntactic or phonological structures of that language, although they are able to use it. The Science that deals with the language is the Linguistics, and the Sciences dealing with the different languages are the different Grammars. The knowledge of these Sciences calls for a big effort and demands a specialization, which gives to the linguists the ability of *speaking with authority* about the language or the different languages, but not about other issues.

According to that, we could define a scientist as the speaker of a language with a *linguistic competence* in order to express himself and develop his profession and his life into the strictly fenced field of his scientific community. Whenever the scientist goes out of this field, he loses his specific competence and, therefore, he is not qualified to speak with authority about other issues different to his own field of knowledge.

When a scientist speaks about Science, the research or the research or teaching institutions, he has not more authority than any other scientist. Moreover, we could say that he has not more authority than other citizens not scientist. A scientist, when speaking about Science or research, *is only expressing his opinions.*

They could speak with authority about Science or the scientific research only if there were a *Science of the Science* or, just the same thing, a metalanguage to analyse every language. In that case, the role would be represented by the *Philosophy* or the *Philosophy of Science.* But this is not the case, because the Philosophy of Science, although developed by an academic community, is not able to create such a consensus to make an agreement about the object and the method among the practising ones.

In the Philosophy of Science there are schools of thought linked to the metaphysical, political or even moral assumptions of its developers, for which reason we can say that, although the philosophers of Science have much more *authority* than the scientists to speak in general terms about Science, they have two serious limitations: first, they don't practice any Science, and because of that, sometimes they ignore the specific process of the research; and secondly, they cannot create a consensus among their schools and tendencies, and so, they are also expressing their opinions.

The fact that the philosophers of Science are not practising scientists and their dependence on the traditionally called Philosophy —namely, its metaphysical or moral assumptions—, leads many of them to develop a certain inferiority complex. This gets them to come closer to the called scientists, to whom many times they offer themselves as an alibi in the strategies developed by the scientific communities in order to take funds and monopolize the academic institutions.

Consequently, we believe that it could be said that nobody has the specific ability to speak in general of Science, the scientific research or the institutions connected with this or with the higher education, which is indispensable for the training of scientists. On this overall level there are only *opinions*. The opinions are formulated from different places, but chiefly from two: from inside the scientific system and from outside the scientific system. Taking into account the general characteristics of the human groups, we could suppose that a fellow pertaining to a group will develop an opinion in favour of his own group in order to get funds or institutional power, and so, he will limit himself to defend the interests of his own group. Given that in the academic or research framework the institutional or economical interests cannot be defended openly —that happens only partly in the fields of Politics or Economy—, the scientist will develop arguments *ad hoc* to defend his strategy. Being used to employ the argument of authority —*argumentum ad baculum*—, derived from his own competence as scientist, he will employ that authority —that he has not in this case— for hiding his real interests under the cover of a scientific pseudo-reasoning.

And so, we could finish the analysis of the weakness of A saying that: *nobody can speak with unappealable authority about the Science, the research or the higher* *education institutions.* The persons taking part in them or funding them or the politicians governing in a certain moment can only express opinions.

All of this demonstrates that the discussions about the Science, the research or the higher education institutions are not «scientific» discussions, but ethical and political. And in those discussions there is nobody with an unappealable authority springing from his competence in a fenced and institutionally accepted field of the research.

**Weakness of 2:**

What is usually called knowledge is really a very heterogeneous reality. The fact that a group of subjects are taught or researched in the universities doesn't mean that they must have necessarily a common internal structure. What they share is an institutional framework: they are explained by professors, namely, persons classified socially in that way. Those professors may develop their activity in similar frameworks (universities, institutes of research), have common patterns of social or academic promotion (*curriculum* standard), but their intellectual activity is radically heterogeneous.

It is generally stated *a priori* that everything made in these institutions is *Science*. Of course, who says such a thing cannot establish demarcation criteria between the Science and other kinds of knowledge, and starts from a principle: the knowledge that can be called Science has only an attribute, it is an unappealable knowledge, endorsed by the existence of a scientific community and the institution that receives it.

We will give some examples. The methods of observation, measure and analysis managed by a physicist have nothing no to do with those used by a zoologist, who establish the systematization of the different animal species, or with those used by an anatomist or a histologist. And even less with the way of working of an historian, a philologist or a penologist jurist.

Those methods of observation are not even common into a field as Physics. Who works in the theory of the superstrings knows that his supposed objects are not observable, because they are confined. And it could happen that other physicists despise this kind of physicist, considering him to be too speculative, almost philosophical. The same thing could be said about what a certain class of physicians think about the psychiatrists —or about the strains between biological and psychological psychiatrists—, or about the strains between surgeons and other kind of physicians, due to different notions of the pathological process.

These same differences occur in fields like the Philology, where the level of abstraction of the general Linguistics has nothing to do with the love for the insignificant details shown by the lexicographer or the historian of language.

If we take into account that there is not unity in the methods of observation and analysis into the same field, stating even that Sciences like the Biology work with three systems hardly compatible —the genetic, the evolutionist and the morphological one—, it is clear that the subject will become more complicated when establishing the languages spoken by the scientist.

A theoretical physicist must know branches of Mathematics very specific and very difficult to manage; other physicists don't need them and their level of mathematical formulization is much lesser. The mathematical knowledge of almost all the chemists is very dim, not to mention the biologists, which main mathematical instrument is Statistics and no always.

The use of Mathematics is important in the economical theory, but lesser in the applied Economy. The Sociology uses Mathematics lesser than the Economy —limiting itself to Statistics—; and History, if not is strictly economical History, manages them on a very elementary level and only as a complement of its narrations. The Law doesn't need Mathematics at all, the same that another fields of knowledge. We cannot say anymore, like Galileo Galilei, that we read the book of Nature using the mathematical characters. And we cannot believe either that Science is the only *well done language*, as Condillac said and the neopositivists believed.

There is not the Science, but only the knowledge, by definition heterogeneous in the methods of getting the information, systematizing and expressing it. The different kinds of knowledge cannot be reduced to a unity. Many theories are incompatible among themselves, even into the same field —as we have seen in the case of Biology, and it happens also in Physics, where the unification of the four basic kinds of strength is only a *desideratum*—.

Given this, it is clear that the knowledge and its process of elaboration cannot be formalizable, even lesser quantifiable, and that there is not any parameter of universal validity for measuring the knowledge and establishing economical or political patterns of funding and planning of such a thing on the overall level.

**Weakness of 3 and 4:**

In all the processes of evaluation is established as a fundamental criterion that there is a unit from which could be stated the intellectual capital of a scientist. This unit is usually the paper or article, both published in a magazine or read in a congress.

Those atoms of intellectual capital are arranged according to a series of criteria which could be considered externalist. Namely: according to the ranking of the magazine, the importance of the congress, and even into this we must take into account the institutional ranking: a poster, a paper, a report or a lecture.

If we examine this fact, something attracts widely our attention, and it is that no criterion of the exposed above values the contents of those contributions, but it limits itself to establish a level of *distinction* against the anonymous mass of the investigators as a whole.

We call *distinction* to the sociological fact of establishing some kind of social ranking based on the possession of a certain class of property: material —money or real state— and symbolic —sacred, artistic, intellectual or military prestige—. These symbolic properties shape an *honors system*, which has only validity into the social group that recognizes it. And so, the military honors are only recognized by who believes in them —the military communities—, and the same happens with the religious honors and, in our case, with the scientific ones.

Consequently, we see that the evaluation of the scientific knowledge in itself doesn't exist, because the contents are not valued, but only the external signs. The evaluation of the scientific knowledge is only an inner matter of the scientific communities, through which they define their rankings and establish their positions for developing their strategies of fundraising and of the control of the research and academic institutions. The existence of these depends, as a last resort, on the funds provided by the State or the industries.

Now we must make clear that it is true the premise established by us, according to which the knowledge is not really valued, but the external signs of it, which is a result of an honors system.

A paper or an article —we will take that unit to simplify— is not a unit of knowledge, but a *rhetorical unit*. If we take the article as a unit of measure, we favour a *literary criterion against a scientific criterion.* The «Science» has a rhetoric analysed, among others, by A.G. Gross (1990). An article must have some formal and stylistic conventions. It is usually assumed that one cannot speak in first person, the data must be presented in a certain way, and be analysed similarly, obtaining the conclusions sometimes with the mathematical formalism and other times without it. The authors of an article cannot have a style of their own, but they must use an impersonal language, similar to the any other article. This language will respect all the conventions about bibliography and the seek of it in the publications socially accepted by the community, which are those with a property called *quality*. That property cannot be defined. A publication must contain truthful principles, and for having quality those principles must be admitted by the respective scientific community, and moreover, they must get enough rank of honour. As a matter of fact, that rank of honour is called «quality».

The *rhetorical unit* is confused with *unit of knowledge*, because to a certain extent there is nothing for it, because there are not units of knowledge.

A Science is not a group of atomic principles that can be numbered (Bermejo Barrera, 2006). On the contrary, a Science is a system of connected propositions building a system or a scientific theory, which is usually called paradigm, term too much misused from the work of T.S. Kuhn (1962).

It could be also that a scientific theory depends on other, as for example Genetics on Biochemistry or the theoretical Physics on certain branches of Mathematics, and so, we must value two levels of connection: the level of connection into a theory and the level of connection between two or more theories.

Given that a scientific theory has an architecture, or with the same words, it is an structure, we cannot establish a simple arithmetical criterion to define the elements that make it up, and even lesser to express it with a whole number or to insist on discover definite factors of classification, like the called *factor h*, exposed enthusiastically by Ricardo García, professor of research of Microelectronics in an article published in the newspaper *El País* (21 December 2005). This factor is defined in this way: «The factor h is a number assigned to each scientist and it establishes the number of articles of that author with so many or more quotations than his factor h».

In that definition, used for the scientific assessment in the twentyfirst century (*sic*), there is no doubt about the article as a unit of measure, the number of articles as a whole number and the quality of the article not considered from its contribution to the knowledge, or epistemic contribution, but from the number of quotations.

But, ¿what is a quotation? A quotation is a sign of recognition made by a member of a scientific community Cc (x) to other member of that scientific community, meaning that the first one grants him authority in the statement or statements made by the author of the paper in the text or in its contents. A quotation is a sign of mutual recognition and the establishment of a criterion of reliability shared by two authors, through which the author of the quotation A recognizes to B his authority to tell the truth.

Being a social process, the more persons confirm the recognition, the better. Moreover, it is being used unconsciously the notion of symbolic capital, ignored by the scientist, and people think that gather a certain number of quotations is equal to gathering a certain quantity of euros or dollars.

Saying that the *factor h* is the number of papers of the author with so many or more quotations than his average is only a liking for the symmetry, or even could be called a kind of mathematical pomposity, because it shows nothing. We equally could say that «the most important brick of a house is the brick in contact with more bricks». We don't think that the future of the architecture depends on that «theorem».

A Science depends on the existence of one or several theories (as in the case of the Biology). In a Science there are two kinds of development and progress of the knowledge: the cumulative one, which occurs into the theory itself and could be quantifiable; and the overall one, also known as «scientific revolution», which involves a general reorganization of a field of knowledge. The overall changes are minimal and they cannot be planned, by definition, for the next reasons:

1. A scientist works inside the framework of a theory. That means that he accepts its principles, uses the methods established by his community and gets some results admitted by the group of scientists who share those principles, use the same methods and show them according to the usual conventions, in the approved publications or in the institutional frameworks valued by all of them.

2. A scientific revolution involves to change the principles (all of them or partly) and the methods, to get unexpected results which could be shown in new forms, for example using other mathematical formulisms, and to split up with the ranking of honors of the scientific community, because those who gather the previous scientific capital will lose most of it when they must reformulate all the system of knowledge. By definition, the number of quotations of preceding authors made by the authors of that «revolution» will be minimal and, consequently, it will be also very small the recognition of the established authority, authority that works through a system of quotations, as T. Grafton (1997) said in his study about the footnotes in the field of Humanities.

Furthermore, the authors often quote and are quoted according to definite data and findings, but never when are established general principles. It happens so the paradox of that the fundamental things shouldn't be quoted, because they are accepted as principles. By this reason, the aim of R. García of stating the correlation between the *factor h* and the Nobel Prize verges on the nonsense by reasons of strictly scientific kind and because the award of the Nobel Prize is a social and political process made by a definite institution. It is very important for the scientists only because symbolizes the top of the whole «system of honors».

We think that from above said it is clear that there is not the unit of measurement of the scientific knowledge, because that knowledge is formed by very complex structures and the ranking of those supposed units of measurement —the papers— is only a part of a process of inner organization of certain social groups, the scientific communities. They have almost nothing to do with the ranking of knowledge. And then, ¿Why the scientists use them? According to two kinds of reasons, sociological and psychological ones.

Sociologically it is necessary to use them because all the human communities need establish their own rankings. In the case of the scientists, their academic promotion and the fundraising for the research depend on the existence of rankings. That academic promotion and that fundraising involve to establish the control of the academic and research institutions and also the development of rhetorical strategies to persuade the politicians and the businessmen of the need of making investments for promoting the scientific knowledge.

In those strategies could be used sometimes only economical arguments of business profitability. But those arguments are not always useful. Moreover, putting them into practice in a consistent way will endanger the existence of the academic institutions and reduce the Science and the research to the development of a simple technology in the service of the companies.

The scientists are aware of that danger, and therefore, they combine wisely the *argumentum ad denarium* with the supposed criteria of excellence and quality, which are only the development of a system of honors and, consequently, of a strategy of ranking oriented to the control of the economic sources and the authority over the institutions.

The psychological need has been analysed by Randall Collins (1998). He says that it is necessary to take into account three factors to understand the scientific work:

> 1. Emotional energy, necessary for making constantly the big effort of the intellectual work.
>
> 2. Accumulation of cultural —or scientific— capital.
>
> 3. Institutional recognition.

A scientist must focus his emotional energy in the achievement of his cultural capital and of his recognition. These two achievements are usually interrelated, although sometimes the recognition and the capital don't happen simultaneously. Scientists with less intellectual capital than others could be recognized by social or political reasons. Moreover, the strategies for obtaining these two parameters use to be clearly distinguished. And so, there are scientists with a great managerial capacity and skills in the social relations, which have lesser cultural capital than other socially awkward scientists. Bill Readings (1996) calls the first ones urban or courtiers and the second ones rural or monks. Of course, the problem rises because the courtiers, due to their ability of accumulating resources and controlling the institutions, could corner or let die the monks of starvation, in symbolic terms. The scientific communities have two characteristics that define the living creatures: they reproduce and tend to take up the whole space for the development of life. A living creature may take up that space and get a balance with the environment or, on the contrary, hasten his extinction due to his overdevelopment. In the life it is obtained a balance when there is a complex system of balance among different living creatures, as states S. J. Gould (2002). In the field of Science maybe would be good to get a provisional balance like that occurring with the viruses, funguses and bacteria.

According to the criteria of weakness of the pseudo-axioms 3 and 4, we think that the scientific evaluation could be defined as a social and political process of fundraising and of control of institutions developed by the scientific communities. That process is unavoidable, but it is corrupted because the prevailing communities impose, by economical or political reasons, their evaluation criterion in order to marginalize the others and corner the academic and research field.

However, the weakness doesn't finish here, not only because in these evaluation processes are not considered another forms of exposing the knowledge, like the books —we won't deal with this—, but also because are established the two last pseudo-axioms, totally unwitting.

**Weakness of 5 and 6:**

It is usually accepted without any argument that the average life of an article is five years. In fact, many times, when a scientist must present a *curriculum*, he will limit himself to his publications and activities in the last five years.

This is supposedly the time in which each article is quoted; afterwards, it will fall into oblivion. If we want to understand how works this system, we should to make a correction on one hand, and on the other, to reveal the existence of the two main ideas used, of which the scientific assessor are not aware.

The correction would be the next: the index of impact of an article cannot be considered equal to the number of its quotations by several reasons.

First, because we tend to believe that a quotation is positive by definition. Namely, an article is quoted in order to recognize its validity, and that is not true in all the fields of knowledge. In the fields of the Social and Law Sciences and in the Humanities many quotations are negative. They are criticisms against what is said in the book or the article and, consequently, refusals of recognition, and no the opposite thing. And so, for example, the book more reviewed and quoted in the field of classic studies in USA is Martin Bernal`s book *Black Athena* (Bernal, 1987). There are more than two hundred reviews of this book, but most of them are negative and show the untenableness of his thesis, which intends to split up with the western historiographic tradition by discovering, supposedly, the African origin of the Greek Culture.

The book was very widely received, due to its political implications, but it was almost unanimously rejected by the specialists. Others published books only for criticizing it (Lefkowitz and McLean, 1996).

Introducing this corrective factor —take away the negative quotations from the positive ones— is not always easy, because some parts of a work can be criticized and others accepted. The scientific criticism hardly tells right from wrong, but it gets an intermediate position.

Although this taking away were possible, to measure the truthful impact we should divide the number of quotations by the *corpus* of all the works in a definite field. So, the impact of the article would be:

$$I(a) = N(c) / N(t) \text{ in } C(x)$$

Being N (c) the number of quotations and N (t) in C (x) the total number of works in a definite field.

We find unnecessary to say that a work quoted ten times in a field in which are published one hundred articles per annum has more impact than other with the same number of quotations in a field in which are published ten thousand articles.

However, the truly important thing is not this, but the assessors work with two pairs of parameters not well formulated, namely:

- Singularity *versus* anonymity.
- Language *versus* silence.

The quotations are important, but not because they measure the scientific knowledge. Taking into account the thousands of works published each year in the fields of the big Sciences and, according to our formula, the impact is minimal. They are important because satisfy the psychological need of recognition of the scientists, reward them for the big efforts made in education and work and because they are a socially useful way of establishing social rankings.

Some years ago, a French physicist and great historian of Science, Gaston Bachelard (1951; 1953) showed that the Physics and the Chemistry are more and more an anonymous knowledge. The groups of work are increasingly bigger and the results obtained become a common asset little by little. The scientist makes efforts to leave the anonymity and the silence, to be quoted, and to get into the world of language and stand out as a public figure against the anonymity. He needs do it to feel satisfaction and to have a cultural capital or, in other words, honors and prestige to perform a social role in the framework of his scientific community.

A scientist develops an academic and research career. The scientific careers are very different, as Tony Becher and Paul R. Trowler (1996) have analysed. The number of articles varies; it is higher in the case of the chemists and the physicists and much lesser in the case of the mathematicians —in them it should be distinguished among different fields—. The philosophers, philologists and historians write hardly books. The work and the logic of writing a book are very different from the production of an article.

In the Social Sciences and the Humanities, almost all the works belong to only one author. In the field of Physics and Chemistry almost never; on the contrary, they use to be works of very large groups. A great philosopher or historian must have his own literary style. None natural scientific has it because he doesn't need have it, and the rhetorical conventions of his discipline don't allow either. A philosopher or an historian must read, think, write and develop a work, at least in a certain part of the cases. A natural scientist doesn't read books or long texts, but basically he does research. He doesn't write, but produces articles; he doesn't think —in the sense of questioning the fundamentals of his own discipline—, and he never have work, but only *curriculum.*

There are very different academic patterns in keeping with the various kinds of *curricula*. In the Becher and Trowler`s book could be seen the big differences, not only with regard to the quantities of the publications, but also considering another very important factors, for example the creativity. There are maybe young physicists very creative, but that is impossible in the field of Philosophy —except Schelling—. The great works of Philosophy have been written by mature men, even with more than sixty years, an age in which a physicist is no more productive. We could think that imposing to all of them the same pattern is a result of the ignorance. Nevertheless, given that the academicians and researchers are not, by definition, ignorant; it is really the aim of privileging some kinds of *curricula* over others in the framework of the academic fight for the fundraising and the control of the institutions.

According to the pseudo-axiom 5, an article casts a cone of light due to the language that fills a space and persists in the time, as shows the sketch of the figure 1.

When that cone of light, merely rhetorical, goes out, the scientist falls into oblivion and his contribution to the development of knowledge becomes a common asset.

An article leaves the cone of light of the quotations and falls into oblivion due two reasons: because it is not of interest, it is obsolete and nobody quotes it. Or because what stated in the article is so important that becomes a part of the essential common asset in the definition of a Science. We could systematize the situation establishing three stages:

1. An article is fundamental and has a large number of quotations.
2. The article is not quoted, but the author will be part of the history of the discipline, with a law or theorem from his name: Fermat`s theorem, Maxwell`s equations... The author becomes so a main character in the history of his Science.
3. The discovering is so important that the author is not named, because what he said is accepted as a fact: Gravity Law and not Isaac Newton`s Law.

According to this pattern, we could formulate the **fundamental paradox of the scientific evaluation**: *A scientist is truly important when nobody quotes his works. His* *works are not quoted because the importance of his discoverings is so great that his name is enough to identify them and, in the future, his name will not be necessary either.*

The simplest way of going out the anonymity is accumulating quotations. But, according to our paradox, that formula nullifies itself. One is truly important when it is not yet necessary to quote him. And the definite importance is falling into oblivion because what one has discovered becomes a fact of universal validity.

Finally, and with regard to what could be called «theory of the perishable knowledge», which is created and destroyed in a very short term, we must show two inconsistencies.

The first one is saying that the magical value of the number 5 is not endorsed by any historical study. We have choose the number 5 because it is the half of 10 and we work in a system of base 10, and because it could be good as an approximation. No reason endorses a whole number, instead of a fraction.

This inconsistency is related with other one, expressed in the pseudo-axiom 6, namely, the next:

If an article loses its value after five years, we may generalize saying that the Science destroys its own past and the scientists don't need know the history of their discipline. However, they can forecast its future, which is contradictory.

The scientists don't need know the history of their discipline because they work with a unique theoretical and methodological pattern in which they are trained, being unable of understanding and practising the patterns of the past. Moreover, they think that this fact is justified by the bigger effectiveness of their pattern. Nevertheless, the scientists will think that they can do without the past of their science, but that is not necessarily true, because we have already established in the analysis of weakness 1 that a scientist may speak with authority about the facts studied by his Science, but not about his Science and even less about the Science in general. A mathematician may not know about Genetics, but that doesn't mean that he has not genes.

The scientists are not aware of the role of the past in the Science of nowadays, because the best researches of the ancient scientists become admitted facts. An astronomer doesn't need show that the Earth is not in the middle of the Universe, because that was already shown by Copernico. And a chemist needn't formulate the atomic theory or build the periodic system, because both things were made by their ancestors, etc.

The scientists receive the legacy of the Science in passing, but they are not aware of it, because they don't know that what now are the facts, in other time were only theories and because many times they don't realize the limitations of their knowledge. And so, for example, before the problem of the «dark matter» any person should understand that the knowledge of the present Physics is only a provisional solution and that the Science, as K. S. Popper said, is not more than a «search without end» and not a closed system.

The scientists think that they can forecast the future of their discipline because they are not aware of their limitations. Undoubtedly, they can plan their research in the short term. But, if following the pseudo-axiom 5, the knowledge expires after five years; we will find a difficulty: the results obtained will destroy the value of the criteria for planning the research, and this has not much sense, because the Science would be so the only system that plans its own destruction. We say destruction and not renovation, because for speaking about renovation we must suppose that the Science with more than five years is not destroyed, but it is cumulated, and this contradicts the pseudo-axiom 5.

Moreover, and this is fundamental, a scientist may establish the prediction of a fact (h) when that fact goes into the field of study of his own science. An astronomer may forecast an eclipse and a physician the death of a patient. Those predictions are not always accurate; they use to be probabilistic, and with the development of the theory of chaos sometimes there is not even a clear statistical regularity. The predictions could be either mechanical or probabilistic, but it is obvious that they must be always included into the field of study of a given Science managed by the scientist.

A scientist cannot forecast the future of his own science and even less of the Science in general, because the scientists don't study their science, but a sector of the reality. The Science is not their object. They haven't a method for studying it, and therefore, they cannot forecast its future, not mechanically or probabilistically.

If there were a «science of the Science», its specialists could forecast the future of the sciences or of the Science. But, as there is not an agree about the existence of such a science, we cannot deify the philosophers of the science as platonic kings-philosophers, owners of the knowledge and the government.

The future of the Science is not predictable. The Science can only be planned in short or medium-term. Although the scientists insist that they have such a capacity of prediction, the fact is not true. They need develop that strategy in order to fundraising money for their current research and to control the academic and research institutions, essential requirement for surviving in a world in which the scientific communities compete a lot. Sometimes, unfortunately, we must add the satisfaction of their vanity, of their ambitions of power and even of economical profit.

After stating the weakness of the pseudo-axioms that endorse the processes of scientific evaluation, we think that it could be said that, while such an evaluation is necessary, because it is impossible to fund everything, the patterns of «measurement» should to be changed according to the next ideas:

1. The Science doesn't exist, but different kinds of knowledge.

2. There isn't one, but many ways of producing knowledge.

3. The ways of exposing the knowledge are very different. There isn't a unity of production of the knowledge.

4. There is not only a model of academic *curriculum*, but many.

5. The diversity of *curricula* demands absolutely heterogeneous ways of valuation.

6. Those valuations must determine both the funding of the research and of the researchers.

7. The ranking of sciences may be epistemic, but not political and economical, and the science of most epistemic rank doesn't need more funding due to it.

8. The heterogeneous production of knowledge is related to different systems of values: technical, economical or human —with values as the Health or the moral and political values—.

9. The pyramid of values is not established by the scientists, but in a given historical period the values are arranged in a certain way.

10. In the present historical time, the Science, understood as a provisional search without end, must be subordinated to the values of the Declaration of Human Rights, recognized by the democratic systems.

11. The democratic systems are only an approximation to the shaping of those values, in the same way that the Science is only an approximation to the conquest of knowledge.

## Coda

In the sixteenth century a shoemaker called Jakob Böhme lived in a village of Germany near to the town of Görlitz. Böhme didn't speak Latin and was the firs philosopher that wrote in German. Because of that Hegel named him, proudly, *philosophus teutonicus.*

Jakob strolled around the fields with a physician that spoke Latin, and he answered him the Latin name of the plants, because he thought that there was a strong connection between the «scientific» name of a plant and its nature. In that time the Latin explained everything, nowadays it is the Science.

In the ancient times there were authors that thought that the parts of the things should to match up with the letters of the words for naming them. Our present scientists make something similar when they believe in the omnipotence of the Science and in its capacity for explaining everything, explaining the Science in itself and explaining even themselves. But Jakob Böhme (Reguera, 2003) had a virtue that the scientists have not: the humbleness. Jakob was a mystic that had some intuitions similar to the current Physics, like many others mystics —which has been proved by Fritjof Capra, 1975). He thought that our origin was the void and we should go back to it. The great mathematician Blaise Pascal was afraid of «the eternal silence of the infinite spaces». But Jakob Böhme wasn`t, because he considered that it was our origin. The prestige of the great scientists dies in the silence, when first their quotations expire, then their figure, and finally they disappear under the facts. In that eternal silence of the infinite spaces we will find some day the Einstein`s ironic smile or our humble shoemaker strolling around the fields, and nobody will assess their *curriculum.*

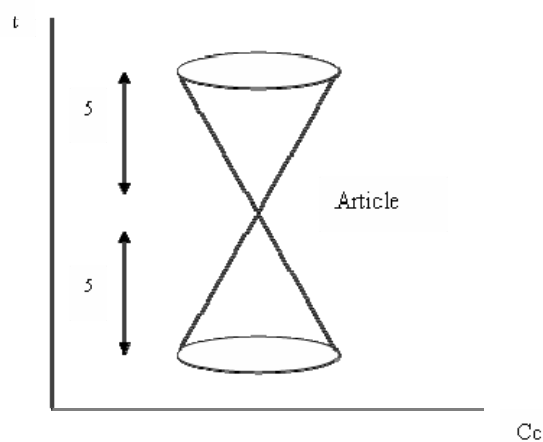

*Figure 1*